\documentclass[prd,preprint,superscriptaddress,nofootinbib,longbibliography]{revtex4-1}
\usepackage{graphicx}
\usepackage{amsmath}
\usepackage{amsfonts}
\usepackage{mathtools}
\usepackage{color}
\usepackage{amssymb}
\usepackage{dcolumn}
\usepackage{bm}
\usepackage{braket}
\usepackage{microtype} 
\usepackage[linktoc=all]{hyperref}
\usepackage{cleveref}
\usepackage{yfonts}
\usepackage{graphicx,epsfig}
\usepackage{epsfig}

\numberwithin{equation}{section}

\newcommand{\ud}{\mathrm{d}}
\newcommand{\op}{\mathcal{O}}
\newcommand{\vevo}{\left\langle\op \right\rangle}
\newcommand{\stresst}{T_{ij}}
\newcommand{\vevt}{\left\langle\stresst \right\rangle}

\newcommand{\be}{\begin{equation}}
\newcommand{\ee}{\end{equation}}
\newcommand{\bea}{\begin{eqnarray}}
\newcommand{\eea}{\end{eqnarray}}
\newcommand{\beas}{\begin{eqnarray*}}
\newcommand{\eeas}{\end{eqnarray*}}
\newcommand{\nn}{\nonumber}
\def\({\left(}
\def\){\right)}
\newcommand{\la}{\langle}
\newcommand{\ra}{\rangle}

\newcommand{\half}{\frac{1}{2}}

\makeatletter
\DeclareRobustCommand{\rcite}[1]{%
\rcite@aux#1,\@nil{#1}%
}
\def\rcite@aux#1,#2\@nil#3{%
\if\relax#2\relax
Ref.~\cite{#3}%
\else
Refs.~\cite{#3}%
\fi
}
\makeatother

\begin{document}
\title{Holographic 2-Point Functions in the Pseudo-Conformal Universe}

\author{Mariana Carrillo Gonz\'alez} 
\email{cmariana@sas.upenn.edu}
\affiliation{Center for Particle Cosmology, Department of Physics and Astronomy,
	University of Pennsylvania, Philadelphia, Pennsylvania 19104, USA}

\author{Kurt Hinterbichler}
\email{kurt.hinterbichler@case.edu}
\affiliation{CERCA, Department of Physics,	\\
	Case Western Reserve University, 10900 Euclid Ave, Cleveland, OH 44106}

\author{James Stokes}
\email{jstokes@flatironinstitute.org}
\affiliation{Center for Computational Quantum Physics and Center for Computational Mathematics, \\
Flatiron Institute, New York, NY 10010 USA}

\author{Mark Trodden}
\email{trodden@physics.upenn.edu}
\affiliation{Center for Particle Cosmology, Department of Physics and Astronomy,
	University of Pennsylvania, Philadelphia, Pennsylvania 19104, USA}

\date{\today}

\begin{abstract}
	\noindent 
	We holographically calculate two-point functions in the pseudo-conformal universe, an early universe alternative to inflation.  The pseudo-conformal universe can be modeled as a defect conformal field theory, where the reheating surface is a codimension-1 spacelike defect which breaks the conformal algebra to a de Sitter subalgebra.  The dual spacetime geometries are domain walls with de-Sitter symmetry in an asymptotically anti-de Sitter spacetime.  We compute 2-point functions of scalars and stress tensors by solving the linearized equations for scalar and tensor fluctuations about these backgrounds.
	
\end{abstract}

\maketitle


\newpage

\section{Introduction}

Alternatives to inflation often involve a pre-big bang phase in which the universe is approximately flat.  Some well studied examples include the ekpyrotic scenario \cite{Khoury:2001wf,Buchbinder:2007ad} and genesis-type models \cite{Creminelli:2006xe,Creminelli:2010ba}.  
Another example, which will be our primary interest, postulates that the early universe is described by a conformal field theory (CFT) on a nearly flat spacetime whose conformal algebra is spontaneously broken by a time-dependent vacuum expectation value (VEV) of the form
\be \la\phi\ra_{\rm }\sim 1/{(-t)}^{\Delta},\label{vevforminte}\ee
where $\phi$ is a dimension $\Delta$ scalar operator.
This VEV breaks the conformal symmetry down to a de Sitter (dS) subalgebra,
\be \mathfrak{so}(4,2)\rightarrow \mathfrak{so}(4,1).\label{symbreakpatternee} \ee
  This is the so-called pseudo-conformal universe \cite{Rubakov:2009np,Creminelli:2010ba,Libanov:2010nk,Hinterbichler:2011qk,Hinterbichler:2012mv,Libanov:2012ev,Libanov:2010ci,Libanov:2015iwa}.   As $t\rightarrow 0$ from below, the VEV \eqref{vevforminte} goes to infinity and the universe must then reheat and transition to the standard big bang radiation dominated phase.  The reheating surface at $t=0$ can be considered as a co-dimension 1 space-like defect in the CFT.  The presence of the defect preserves a dS subgroup and is responsible for the symmetry breaking \eqref{symbreakpatternee}.

Ideally one would like to have complete examples of CFTs which possess states with the required VEVs to realize the pseudo-conformal mechanism. This issue has been addressed in \cite{Hinterbichler:2014tka,Libanov:2014nla} where holographic constructions of the pseudo-conformal mechanism are found using the AdS/CFT \cite{Maldacena:1997re} correspondence.  Another advantage of the holographic approach is that it opens up the possibility for a strongly coupled early universe scenario, in contrast to standard scenarios such as inflation which are typically weakly coupled.

The holographic dual to the four-dimensional pseudo-conformal universe is a five-dimensional asymptotically AdS space in which there is a domain wall that is foliated by slices that are four-dimensional inflationary-patch dS spaces.  In \cite{Hinterbichler:2014tka}, a spacetime background satisfying these requirements was identified in the context of pure Einstein gravity minimally coupled to a massless scalar field, and one-point functions of the fields were computed, verifying the existence of VEVs with the correct symmetry breaking pattern.  

In this paper we compute two-point functions.  These are the observables which are relevant for the computation of power spectra in cosmology.  In the AdS/CFT dictionary, the one-point functions in the dual CFT are determined by the background configurations of fields in the bulk, whereas two-point functions are determined by linear fluctuations on top of the bulk geometry.  Thus, to obtain the two-point functions we must solve the linearized equations for bulk fluctuations on top of the domain wall.  
Our strategy will be to employ a coordinate system in which the spatial slices, and the boundary geometry, are $dS_4$ spaces.  In this slicing, the required VEV \eqref{vevforminte} which breaks conformal symmetry down to dS symmetry becomes simply a constant $\la\phi\ra_{dS}\sim {\rm const.}$  Once the correlators on $dS_4$ are obtained, they can be Weyl transformed to recover the original flat space correlators of interest.

\section{Holographic CFT Correlators}

We start by reviewing the formalism that relates CFT correlation functions to bulk fields.
The AdS/CFT correspondence tells us that for every bulk field $\phi$ there is a corresponding single-trace operator ${\cal O}$ of the large $N$ CFT which lives at the boundary.  The standard AdS/CFT prescription \cite{Witten:1998qj,Gubser:1998bc} tells us that the generating function of the correlation functions of the operators ${\cal O}$ of the CFT defined on the boundary $\partial \mathcal{M}$ of the bulk spacetime $\mathcal{M}$,
\begin{equation}
Z_{\text{CFT}}[\Phi_0]\equiv\left\langle e^{-      \int_{\partial \mathcal{M}}\Phi_0 \op} \right\rangle \ , \label{adscft}
\end{equation}
is given at leading order in large $N$ by the extrema of the bulk gravitational action on $\mathcal{M}$,
\begin{equation}
Z_{\text{CFT}}[\Phi_0]= e^{-S_{\text{on-shell}}[\Phi_0]} \ . \label{adscft2}
\end{equation}
Here $\Phi_0$ are the boundary values of the bulk fields $\Phi$, which act as sources for the dual operators in the CFT.   At tree-level, which corresponds to leading order in $1/N$, the bulk action is to be evaluated for the on-shell solution that reduces to $\Phi_0$ on the boundary.

More precisely, the bulk field $\Phi$ satisfies the boundary condition
\begin{equation}
\Phi_0(x)=\lim\limits_{z\rightarrow 0}z^{\Delta-4}\Phi(x,z) \ ,
\end{equation}
where $\Delta$ is the scaling dimension of $\op$. 
In this expression, we are using Fefferman-Graham coordinates \cite{AST_1985__S131__95_0} in which the metric is written as
\begin{equation}
d s^{2}=\frac{L^2}{z^{2}}\left(d z^{2}+g_{ij}(x, z) d x^{i} d x^{j}\right), \ \ \ z>0,\ \ \ \label{FG}
\end{equation}
covering the Poincar\'e patch in the AdS case for which $g_{ij}=\eta_{ij}$ and $L$ is the AdS radius.  In these coordinates, the boundary is located at $z\rightarrow 0$. 

We will be interested in a bulk theory corresponding to a  scalar field living in a curved background; that is, the bulk fields are a graviton and a scalar. Near the boundary, these fields have an asymptotic expansion of the form
\begin{align}
g_{ij}(x, z)&={g_{(0)}}_{ij}(x)+z^2{g_{(2)}}_{ij}(x)+z^{4}\left({h_{(0)}}_{ij}(x)+{g_{(4)}}_{ij} (x)\log{z}+\cdots\right)+\cdots \ , 
\end{align}
\bea \Phi(z,x)=&& z^{4-\Delta}\left[\phi_{(0)}(x)+z^2\phi_{(2)}(x)+z^4\phi_{(4)}(x)+\cdots\right] \nn \\ 
&& +z^{\Delta}\left[\varphi_{(0)}(x)+z^2\varphi_{(2)}(x)+z^4\varphi_{(4)}(x)+\cdots\right],\ \ \ \ \Delta\notin {\rm Integer}.\eea
\bea \Phi(z,x)=&&z^{4-\Delta}\bigg[\phi_{(0)}(x)+z^2\phi_{(2)}(x)+\cdots+z^{2\Delta-6}\phi_{(2\Delta-6)}(x)+z^{2\Delta-4}\left(\varphi_{(0)}(x)+\phi_{(2\Delta-4)}(x)\ln z\right)\nn\\
&&+z^{2\Delta-2}\left(\varphi_{(2)}(x)+\phi_{(2\Delta-2)}(x)\ln z\right)+\cdots\bigg],\ \ \ \Delta\in {\rm Integer} \ ,\eea
where $\Delta=2+ \sqrt{{4}+m^2L^2}$ is the scaling dimension of the dual CFT operator.  

The $4D$ fields $\phi_{(0)}$ and $\varphi_{(0)}$ can be considered as the two independent boundary data for the $5D$ bulk equations of motion.  Using the bulk equations of motion, the $\phi_{(n)}$, $n\geq 2$ are determined in terms of $\phi_{(0)}$, and the $\varphi_{(n)}$, $n\geq 2$ are determined in terms of $\varphi_{(0)}$.  The function $\phi_{(0)}$ is the boundary value corresponding to the source for a scalar operator in the dual CFT,  
and $\varphi_{(0)}$ turns out to correspond to the VEV of that operator, with its value determined by additional boundary conditions deep in the bulk.   

Similarly, the Einstein equations will require two pieces of boundary data. One of these is the boundary metric
$g_{(0)ij}$, and a near boundary expansion of the equations of motion will determine the $g_{(k)ij}$ for $k\geq2$ in terms of $g_{(0)ij}$.  
The other piece of boundary data is $h_{(0)ij}$, with the $h_{(k)ij}$ for $k\geq2$ determined in terms of $g_{(0)ij}$ and $h_{(0)ij}$. However, we are not completely free to choose $h_{(0)ij}$; its trace and divergence (w.r.t. $g_{(0)ij}$) will be determined, and the rest of $h_{(0)ij}$ will be set by boundary conditions other than Dirichlet data at $z=0$ (i.e. data at $z=\infty$).

The on-shell action is generally divergent and a proper renormalization procedure must be used \cite{deHaro:2000vlm,Skenderis:2002wp}. The renormalized action is defined as
\begin{equation}
S_\text{ren}[\Phi_0]=\lim_{\epsilon\rightarrow 0}\left(S_\text{reg}[\Phi_0;\epsilon]+S_\text{ct}[\epsilon]\right) \ ,
\end{equation}
where $S_\text{reg}$ is the on-shell action with the radial integration domain restricted to $z>\epsilon$ and $S_\text{ct}$ is a counterterm action consisting of purely local terms on the boundary, chosen to cancel the divergent terms in the regularized action $S_\text{reg}$ as $\epsilon\rightarrow 0$. The regularized action and the corresponding counterterms that lead to the renormalized action for the proposed dual of the Pseudo-Conformal Universe can be found in \cite{Hinterbichler:2014tka}. After renormalizing the action, the resulting one-point functions are
\begin{align}
\left\langle\op(x)\right\rangle_s&=\frac{1}{\sqrt{|g_{(0)}(x)|}}\frac{\delta S_{\text{ren}}}{\delta \phi_{(0)} (x)} \ ,\nn \\ 
\left\langle\stresst \right\rangle_s&=-\frac{2}{\sqrt{|g_{(0)}(x)|}}\frac{\delta S_{\text{ren}}}{\delta {g_{ij}}_{(0)} (x)} \ , \label{onepointsfe}
\end{align}
where $S_{\text{ren}}$ is the on-shell renormalized action, and the subscript $s$ denotes the correlation functions in the presence of sources.  The higher point correlation functions can then be obtained with further functional derivatives of the one-point functions with respect to the sources.

In the CFT, there are Ward identities corresponding to global symmetries. In AdS/CFT, there exists a  correspondence between gauge symmetries of the bulk (in our case only diffeomorphisms) and global symmetries of the boundary theory.  We can understand the Ward identities of the CFT correlators by performing shifts on the action given by bulk diffeomorphisms. Consider the variation of the renormalized action, which upon using \eqref{onepointsfe} is given by
\begin{equation}
\delta S_\text{ren}=\int\ud^4x\sqrt{|g_{(0)}|}\left(-\frac{1}{2}\vevt_s \delta g_{(0)}^{ij}+ \vevo_s \delta \phi_{(0)} \right). \label{sren}
\end{equation}
The Ward identities correspond to diffeomorphisms that leave the Fefferman-Graham form of the metric invariant. The infinitesimal action of these symmetries on the sources is given by
\begin{align}
\text{boundary diffeomorphisms:}\quad & \delta g_{(0)}^{i j}=\left(\nabla^{i} \xi^{j}+\nabla^{j} \xi^{i}\right), \quad \delta \phi_{(0)}=\xi^{i} \nabla_{i} \phi_{(0)}   \ , \\
\text{boundary Weyl transformations:}\quad & \delta g_{(0)}^{i j}=2 \sigma g_{(0)}^{i j}, \quad \delta \phi_{(0)}=-(4-\Delta) \sigma \phi_{(0)}  \ ,
\end{align}
which correspond to diffeomorphisms transverse to the radial coordinate and a subset of the $5D$ diffeomorphisms that act as a Weyl transformation on the boundary metric.  These symmetries imply the Ward identities
\begin{subequations}
	\begin{align}
	\nabla^{i}\left\langle T_{i j}\right\rangle_{s}&=-\langle O\rangle_{s} \nabla_{j} \phi_{(0)} \ , \\
	\left\langle T_{i}^{i}\right\rangle_{s}&=-(4-\Delta) \phi_{(0)}\langle O\rangle_{s}+\mathcal{A} \ , \label{tracet}
	\end{align}
	\label{ward}
\end{subequations}
where $\mathcal{A} $ is the Weyl anomaly that arises because the regularized action necessarily breaks the radial diffeomorphism symmetry that induces the Weyl transformation \cite{Henningson:1998gx}.

\section{VEV deformations and two-point functions}
We will be interested in the case of  a CFT where the scalar develops a VEV.  In this case, the source field is zero and the leading term in the asymptotic expansion is the $z^\Delta$ term.  The VEV turns out to be proportional to $\varphi_{(0)}$, and since the VEV is non-zero, this will break conformal invariance.  For the Pseudo-Conformal Universe dual, we require a time dependent VEV of the form \eqref{vevforminte}, which realizes the symmetry breaking pattern \eqref{symbreakpatternee}. 

If we use Fefferman-Graham coordinates in which the slices of constant radial coordinate are curved $dS_4$ slices, then the VEV will appear to be coordinate independent \cite{Hinterbichler:2014tka}.   
Our approach thus consists of first computing the correlation functions for a CFT living in a curved slicing with $dS_4$ metric $\gamma_{ij}$ and afterwards performing a Weyl transformation to obtain the correlation functions for the dual CFT living in flat space.  CFT correlation functions on spaces of constant curvature have been studied previously, see e.g. \cite{Osborn:1999az,Alvarez:2020pxc} (see also \cite{Hinterbichler:2015pta} for other details on holographic CFTs on maximally symmetric spacetimes).

While the expressions for the one-point functions can be obtained through a near-boundary analysis, higher order correlators require taking into account the dynamics of fluctuations inside the bulk. In most cases, an exact solution of the non-linear equations is not available. However, the two-point functions only require the linear dependence on the sources, and thus we can compute the two-point functions by solving for the linearized fluctuations around the background \cite{Skenderis:2002wp}.

Consider  an  asymptotically AdS  ``domain wall''
background of the form 
\begin{equation}
 \bar{g}_{\mu \nu} d x^{\mu} d x^{\nu}=n(\rho)^{2} d \rho^{2}+a(\rho)^{2} \gamma_{i j} d x^{i} d x^{j}, \quad \bar{\phi}=\bar{\phi}(\rho) \ , \label{domainwall}
\end{equation}
where $\rho$ is some general radial coordinate of the bulk and $\gamma_{i j} $ is a maximally symmetric domain wall metric with curvature $k$  normalized as 
\begin{equation}
R(\gamma)_{i j k l}=k\left(\gamma_{i k} \gamma_{j l}-\gamma_{j k} \gamma_{i l}\right) \ .
\end{equation}
In a cosmological setting, $n$   would be the lapse function and $a$ the scale factor.  

We define the perturbations of the metric and scalar field respectively as $h_{\mu\nu}$ and $\varphi$, i.e.
\begin{equation}
g_{\mu\nu}=\bar{g}_{\mu\nu}+h_{\mu\nu} \ , \quad \phi=\bar{\phi}+\varphi \ . \label{pert1}
\end{equation}
All the following expressions are $\gamma$-covariant; that is, indices are raised and lowered by  $\gamma_{i j} $ and the covariant derivatives and curvature tensors are those of the $\gamma$ metric.  Analogously to cosmological perturbation theory (see e.g. \cite{Weinberg:2008zzc}), the metric perturbation can be decomposed as
\begin{align}
h_{\rho  \rho}&= n^{2} \Phi  \ , \nonumber \\
h_{i  \rho}&=n a\left(\nabla_{i} B+v_{i}\right) \ , \nonumber \\
h_{i j}&=a^{2}(\rho)\left(\zeta \delta_{i j}+\nabla_{i} \nabla_{j} \chi+\nabla_{i} w_{j}+\nabla_{j} w_{i}+h_{i j}^{T T}\right) \ , \label{pert}
\end{align}
where
\begin{equation}
\nabla^{i} v_{i}=0, \quad \nabla^{i} w_{i}=0, \quad \nabla^{i} h_{ij}^{T T}=0, \quad \gamma^{i j} h_{i j}^{T T}=0 \ .
\end{equation}

We now insert this into the expression \eqref{sren} for the variation of the renormalized action, and keep terms only to linear order in these perturbations.   After using the Ward identities, Eq.\eqref{ward}, and assuming that the background VEV or the source are constant, we find\footnote{Notice that the one-point functions depend on  the sources but we have removed the subscript $s$ for simplicity.}
\begin{align}
\delta S_\text{ren}&=\int\ud^4 x \sqrt{|\gamma|} \left(1+2\zeta_{(0)}+\frac{1}{2}\nabla^2\chi_{(0)}\right)\Bigg(-\frac{1}{2}\vevt\  {\delta h^{T T}_{(0)}}^{ \ ij}-\frac{1}{2}\langle{T_i}^i\rangle \delta\zeta_{(0)}  \nonumber \\
& +\frac{1}{2}\vevo(\nabla^2\varphi_{(0)})\delta\chi_{(0)}+\vevo \delta\varphi_{(0)} \Bigg) \ . \label{renx}
\end{align}

Our next step is to transform to Fourier space. To do so, we use a basis of eigenfunctions of the Laplacian of the metric $\gamma$. In flat space this basis simply corresponds to plane waves, but here we need to consider the basis appropriate for the symmetries of the curved slicing.  We will not need to be explicit about the basis --- it will be enough to know that it is complete and orthonormal.   Define $Y_\lambda(x)$ to be a complete set of orthonormal eigenfunctions of the scalar Laplacian with eigenvalues $\lambda$,
\begin{equation}
-\nabla^2 Y_\lambda(x)= \lambda Y_\lambda(x) \ .
\end{equation} 
The completeness and orthonormality relations are
\begin{align}
\int d^{4} x \sqrt{|\gamma|} Y_{\lambda}(x) Y_{\lambda^{\prime}}(x)&=\delta_{\lambda \lambda^{\prime}}\ , \\
\sum_{\lambda} Y_{\lambda}(x) Y_{\lambda}\left(x^{\prime}\right)&=\frac{1}{\sqrt{|\gamma|}} \delta^{4}\left(x-x^{\prime}\right) \ .
\end{align}
Given this, any function $F(x)$ can be expanded over the basis $Y_\lambda$ to yield an associated Fourier transform $\tilde F(\lambda)$, with the Fourier transform pair given by 
\begin{equation}
F(x)=\sum_{\lambda} \tilde{F}(\lambda) Y_{\lambda}(x), \quad \tilde{F}(\lambda)=\int d^{4} x \sqrt{|\gamma|} F(x) Y_{\lambda}(x) \ .
\end{equation}

Using this, we can write the variation of the renormalized action \eqref{renx} in Fourier-space as 
\begin{align}
\delta S_\text{ren}&=\sum_{\lambda} \Big(-\frac{1}{2}\left(\vevt_B+\vevt_\delta\right){\delta h^{T T}_{(0)}}^{ \ ij}-\frac{1}{2}\left(\langle{T_i}^i\rangle_B+\langle{T_i}^i\rangle_\delta \right)\delta\zeta_{(0)} -\frac{\lambda}{2}\vevo_B \varphi_{(0)}  \delta\chi_{(0)} \nonumber \\
&+ \vevo_\delta \delta\varphi_{(0)} + 2\vevo_B \zeta_{(0)} \delta \varphi_{(0)}-\frac{\lambda}{2}\vevo_B  \chi_{(0)} \delta\varphi_{(0)}  \Big) \ ,  \label{renp}
\end{align}
where we have divided the scalar and tensor VEVs into the background and perturbation contributions as $\vevo=\vevo_B+\vevo_\delta$ and $\vevt=\vevt_B+\vevt_\delta$. Every term in this expression depends on $\lambda$ but we have suppressed this dependence. We have also omitted any terms that do not contribute to the 2-point functions in this expression and we will keep doing so for all other expressions that follow. Using this result, we can write the non-vanishing two-point functions as:
\begin{align}
\left\langle\op(\lambda)\op(\lambda)\right\rangle&\equiv - \frac{\delta^2 S}{\delta\varphi_{(0)}(\lambda)\delta \varphi_{(0)}(\lambda)}=- \frac{\delta \vevo}{\delta  \varphi_{(0)}} \ ,\\
\left\langle\stresst^{TT}(\lambda)T_{kl}^{TT}(\lambda)\right\rangle&\equiv- 4  \frac{\delta^2 S}{\delta {h^{TT}_{(0)}}^{ \ ij}(\lambda)\delta {h^{TT}_{(0)}}^{ \ kl}(\lambda)}=2\frac{\delta \left\langle T_{ij} \right\rangle}{\ \delta  {h^{TT}_{(0)}}^{ \ kl}} \ ,\\
\left\langle T_i^i(\lambda)\op(\lambda)\right\rangle& \equiv 2 \frac{\delta^2 S}{\delta\zeta_{(0)}(\lambda)\delta \varphi_{(0)}(\lambda)}=4 \vevo_B+2 \frac{\delta \vevo}{\delta  \zeta_{(0)}}  \ , \label{to}\\
\nabla^i\nabla^j\left\langle\stresst(\lambda)\op(\lambda)\right\rangle&\equiv 2   \frac{\delta^2 S}{\delta\chi_{(0)}(\lambda)\delta \varphi_{(0)}(\lambda)}= - \lambda \vevo_B \ .
\end{align}
Using the Ward Identities and Eq. \eqref{to}, we can see that 
\begin{equation}
\frac{\delta \vevo(\lambda)}{\delta  \zeta_{(0)}(\lambda')}=-\frac{\Delta}{2} \langle O\rangle_{B}\delta_{\lambda \lambda'}\ .\label{fromwi}
\end{equation}

Now, transforming back to coordinate space, we find that the position-space 2-point functions are given by, e.g.
\begin{equation}
\left\langle \op(x)\op^{\prime}(x^{\prime})\right\rangle = \sum_{\lambda} Y_{\lambda}(x) Y_{\lambda}\left(x^{\prime}\right) \ \left\langle\op(\lambda)\op^{\prime}(\lambda)\right\rangle \ .
\end{equation}

The last step is to perform a Weyl transformation to find the 2-point functions of the flat space CFT. 
In our case, since maximally symmetric spaces are conformally flat, the flat boundary metric is related to $\gamma$ through $\eta_{\mu \nu}(x)=\Omega^{2}(x) \gamma_{\mu \nu}(x) \ $, with the Weyl factor $\Omega>0$; thus, we can obtain the correlation functions on the AdS boundary from the correlation functions computed in the curved slicing as
 \begin{align}
 \left\langle\mathcal{O}_{1}\left(x_{1}\right) \cdots \mathcal{O}_{2}\left(x_{n}\right)\right\rangle_{\eta}=&\Omega^{-\Delta_{\mathcal{O}_{1}}}\left(x_{1}\right) \cdots \Omega^{-\Delta_{\mathcal{O}_{n}}}\left(x_{n}\right)\Bigg(\left\langle\mathcal{O}\left(x_{1}\right) \cdots \mathcal{O}\left(x_{n}\right)\right\rangle_{\gamma} \nonumber \\
&+(-1)^{n+1}\frac{1}{\sqrt{|g_{(0)}|}}\frac{\delta^n S_\mathcal{A}}{\delta \phi_{(0)} (x_1)\cdots\delta \phi_{(0)} (x_n)} \Big|_{\phi_{(0)}=0} \Bigg) \ , \label{Weylt}
 \end{align}
where $\Delta_{\mathcal{O}_{i}}$ is the scaling dimension of $\mathcal{O}_{i}$.   The term $S_\mathcal{A}$ is a contribution from the Weyl anomaly \cite{Henningson:1998gx}; in general it is theory dependent, but it is always local and so does not contribute at separated points.  We will not keep track of it in what follows.

The desired CFT 2-point functions are given by the following
 \begin{align}
 \left\langle\op(x)\op(x^{\prime})\right\rangle_{\eta}&=-\Omega^{-\Delta}(x)\Omega^{-\Delta}(x^\prime)\sum_{\lambda} Y_{\lambda}(x) Y_{\lambda}\left(x^{\prime}\right) \  \frac{\delta \vevo}{\delta  \varphi_{(0)}} \ ,\\
 \left\langle\stresst^{TT}(x)T_{kl}^{TT}(x^{\prime})\right\rangle_{\eta}&=2\ \Omega^{-4}(x)\Omega^{-4}(x^\prime)\sum_{\lambda} Y_{\lambda}(x) Y_{\lambda}\left(x^{\prime}\right) \frac{\delta \left\langle T_{ij} \right\rangle}{\ \delta  {h^{TT}_{(0)}}^{ \ kl}} \ ,
 \end{align}
accompanied by two other expressions that are fixed by Ward identities 
\begin{align}
 \left\langle T_i^i(x)\op(x^{\prime})\right\rangle_{\eta}&= \Omega^{-4}(x)\Omega^{-\Delta}(x^\prime) \left( 4 \vevo_B \frac{1}{\sqrt{|\gamma|}} \delta^{4}\left(x-x^{\prime}\right)  +\sum_{\lambda} Y_{\lambda}(x) Y_{\lambda}\left(x^{\prime}\right) 2 \frac{\delta \vevo}{\delta  \zeta_{(0)}} \right)  \ , \nonumber \\
& = \Omega^{-4}(x)\Omega^{-\Delta}(x^\prime)  \left(4-\Delta\right) \vevo_B \frac{1}{\sqrt{|\gamma|}} \delta^{4}\left(x-x^{\prime}\right)   \ ,\label{lasttwoeq1}\\
\nabla^i\nabla^j \left\langle\stresst(x)\op(x^{\prime})\right\rangle_{\eta}&= -  \Omega^{-4}(x)\Omega^{-\Delta}(x^\prime)  \vevo_B \sum_{\lambda} Y_{\lambda}(x) Y_{\lambda}\left(x^{\prime}\right)  \lambda\  , \nonumber \\
& =  \Omega^{-4}(x)\Omega^{-\Delta}(x^\prime)  \vevo_B \nabla^2\left(\frac{1}{\sqrt{|\gamma|}} \delta^{4}\left(x-x^{\prime}\right) \right) \  , \label{lasttwoeq2}
 \end{align}
where in both cases the second equality was obtained by using Eq.\eqref{ward}. Notice that the correlators \eqref{lasttwoeq1}, \eqref{lasttwoeq2} are manifestly local, which is consistent with the conservation and tracelessness of the stress-energy tensor at separate points.

\section{Gauge invariant perturbations }
We now turn to solving the bulk equations of motion of the linearized perturbations.  We first must isolate the bulk gauge-invariant perturbations. The gauge freedom corresponds to diffeomorphisms which at the linearized level act on the perturbations as
\begin{equation}
\delta h_{\mu \nu}=\nabla_{\mu} \xi_{\nu}+\nabla_{\mu} \xi_{\nu}, \quad \delta \varphi=\xi^{\mu} \partial_{\mu} \phi \  ,  \label{lingauge}
\end{equation}
with  $\xi_{\nu}$ an arbitrary vector  gauge parameter. These diffeomorphisms are a symmetry of the quadratic action for the perturbations. Similar to our decomposition of the perturbation in Eq.\eqref{pert}, we decompose the gauge parameter $\xi_\mu$ as 
\be \xi_i=\nabla_i\epsilon^S+\epsilon_i^V,\ \ \ \xi_0=\epsilon_0,\ee
where
\be \nabla^i\epsilon_i^V=0.\label{TT2}\ee

Expanding out the linearized gauge transformations \eqref{lingauge} we can find then the following gauge transformation rules for the perturbations of the metric and scalar field 
\bea 
\delta \Phi&=&{2\over n}{d\over d\rho}\left({1\over n}\epsilon_0\right), \nonumber\\
\delta B&=&{1\over an}\left(\epsilon_0-2{a'\over a}\epsilon^S+ \epsilon'^S\right),\nonumber\\
\delta \zeta&=&{2\over n^2}{a'\over a}\epsilon_0,\nonumber\\
\delta \chi&=&{2\over a^2}\epsilon^S,\nonumber\\
\delta \varphi &=& {{\phi'}\over n^2}\epsilon_0, \nonumber\\ 
\delta v_i&=&{1\over an}\left( {\epsilon'}_i^V-2{a'\over a}\epsilon_i^V\right),\nonumber\\
\delta w_i&=&{1\over a^2}\epsilon_i^V,\nonumber\\ 
\delta h^{TT}_{ij}&=&0 \ ,
\eea
where a prime denotes a derivative with respect to the radial coordinate $\rho$. Note that the transverse traceless tensor modes $h^{TT}_{ij}$ are automatically gauge invariant. From these expressions, we can construct gauge invariant variables for the perturbations which are given by
\bea && \tilde v_i=v_i-{a\over n} w'_i,\ \ \ \ \tilde \zeta=\zeta-{2a'\over n}B+{aa'\over n^2} \chi',\ \ \  \tilde \Phi=\Phi-{2\over n}{d\over d\rho}\left(aB\right)+{1\over n}{d\over d\rho}\left({a^2\over n} \chi'\right), \nonumber\\ 
&& \tilde \varphi=\varphi-{a\over n}{\phi'} \left( B-{1\over 2} {a\over n}  \chi'\right).\label{gaugecombsb}
\eea
These are the analog of the Bardeen variables \cite{Bardeen:1980kt} in cosmology and will be the variables used to solve the equations of motion.

\subsection{The quadratic action}

With gauge invariant perturbation variables in hand, we turn to constructing the corresponding quadratic action. As we have mentioned, our setup consists of a domain wall spacetime sourced by a scalar field.  The Lagrangian is that of a canonical scalar with potential $V(\phi)$ minimally coupled to gravity in $5$ dimensions 
\begin{equation}
 \mathcal{L}=  \sqrt{-g}\left[{1\over 2} R-{1\over 2}\(\partial\phi\)^2-V(\phi)\right],  \label{action}
\end{equation}
where the overall $5D$ Planck mass has been set to unity.  The equation of motion for the scalar is
\be \square\phi-V'(\phi)=0,\label{scalareq} \ee
and the Einstein equations for the metric are
\be R_{\mu\nu}-{1\over 2}R g_{\mu\nu}=\nabla_\mu\nabla_\nu\phi-{1\over 2}g_{\mu\nu}\(\partial\phi\)^2-g_{\mu\nu}V(\phi)\ \label{einsteineq}.\ee 
By considering the perturbations around an arbitrary background solution of these equations, as in Eq.\eqref{pert1}, we  find the quadratic Lagrangian is given by
\bea {1\over \sqrt{-g}}{\cal L}^{(2)} = &&-{1\over 8}\nabla_\alpha h_{\mu\nu} \nabla^\alpha h^{\mu\nu}+{1\over 4}\nabla_\alpha h_{\mu\nu} \nabla^\nu h^{\mu\alpha}-{1\over 4}\nabla_\mu h\nabla_\nu h^{\mu\nu}+{1\over 8} \nabla_\mu h\nabla^\mu h \nonumber\\
&& +{1\over 4}V(\phi)\left( h^{\mu\nu}h_{\mu\nu}-\half h^2\right) +h^{\mu\nu}\partial_\mu\phi\partial_\nu\varphi-{1\over 2}h\left(\partial_\mu\phi\partial^\mu\varphi+V'(\phi)\varphi\right) \nonumber\\
&&-{1\over 2}\left[(\partial\varphi)^2+V''(\phi)\varphi^2\right] \ ,  \label{quadaction}
\eea
where everything here is covariant with respect to the background metric $\bar{g}_{\mu\nu}$. This expression was obtained after integrating by parts and  using the background equations of motion.

Now, we specialize to the case of a domain wall background as in Eq.\eqref{domainwall}.  With this ansatz, the two independent equations of motion are the scalar equation of motion, which becomes
\be \phi''+\left(4 {a'\over a} -{n'\over n}\right)\phi'-n^2V'(\phi)=0 \ ,\label{FRWback1}\ee
and the $00$ component of the Einstein equation, which becomes
\be {6}\left({a'^2\over a^2}-k{n^2\over a^2}\right)={1\over 2}\phi'^2-n^2V(\phi) \ .\label{FRWback2}\ee
We can further decompose the Lagrangian in terms of the gauge invariant variables in Eq.\eqref{gaugecombsb}. In the final expressions the tensor, vector, and scalar sectors decouple and their Lagrangians read
\begin{itemize}
	\item   tensor sector:
	\be {1\over \sqrt{|\gamma|}}{\cal L}_{\rm tensor}={na^4\over 8}\left[{1\over n^2} h_{ij}'^{TT\ 2}+{1\over a^2}h_{ij}^{TT}\left(\Delta_L+6k\right)h^{TT ij}\right],\label{tensorexp}\ee
	\item   vector sector:
	\be {1\over \sqrt{|\gamma|}}{\cal L}_{\rm vector}={na^{2}\over 4}  \tilde v_{i}\left(\Delta_L+6k\right)\tilde v ^{i} ,\label{vector}\ee
	\item  scalar sector:
	\bea {1\over \sqrt{|\gamma|}}{\cal L}_{\rm scalar}=&&na^4\bigg[ {3\over 2n^2}\( {\tilde\zeta'}-{a'\over a}\tilde\Phi\)^2-{3\over 4a^2}\tilde\zeta\( \Delta_L+4k\)\tilde\zeta \nonumber\\ 
	&&-{3\over 4a^2}\tilde\zeta \(\Delta_L+4k\)\tilde\Phi -{\phi'^2\over 8n^2} \tilde \Phi^2+{1\over 2}\left(-{1\over n^2} {\tilde{\varphi'}}^2+{1\over a^2}\tilde\varphi\Delta_L\tilde\varphi-\partial^2_\phi V \ \tilde\varphi^2\right) \nonumber\\
	&& -{1\over 2}\partial_\phi V   \ \tilde\varphi\(\tilde\Phi+4\,\tilde\zeta\)+{\phi '\over 2n^2}{\tilde\varphi'}\(\tilde\Phi-4\,\tilde\zeta\)\bigg] .\label{scalar}\eea
\end{itemize}
Here, $\Delta_L$ is the Lichnerowicz Laplacian, which acting on a rank-$s$ spatial tensor is
\be \Delta_L =\nabla_i\nabla^i-ks(s+2).\ee
This is the natural Laplacian to use on a maximally symmetric space, because it commutes with covariant derivatives and with traces. When obtaining the expressions for the Lagrangians above, we have again made heavy use of integration by parts, as well as the background equations Eq. \eqref{FRWback1} and Eq. \eqref{FRWback2}.

Focusing on the scalar sector, we notice that the $\tilde{\Phi}$ variable appears in the action without any time derivatives, and so we can eliminate it as an auxiliary variable,
\begin{equation}
\tilde\Phi
= \frac{1}{12H^2-\phi'^2}\left[12H\tilde\zeta' + 3(n/a)^2(\Delta_L + 4k)\tilde\zeta - 2\phi'\varphi'\right] \ .
\end{equation}
Substituting back into the action, and considering the case of a constant potential which is the case of current interest, we find
\begin{equation}
\mathcal{L}_{\rm scalar}
= \mathcal{L}_{\tilde\varphi} + \mathcal{L}_{\tilde\zeta} + \mathcal{L}_{\tilde\varphi\tilde\zeta}
\end{equation}
where
\begin{align}
\frac{1}{\sqrt{|\gamma|}}\mathcal{L}_{\tilde\varphi}
& = na^4 \bigg[\frac{1}{2a^2}\tilde\varphi \Delta_L \tilde\varphi - \frac{6H^2}{n^2\big[12H^2 - \phi'^2\big]}{\tilde\varphi}'^2\bigg], \\
\frac{1}{\sqrt{|\gamma|}}\mathcal{L}_{\tilde\zeta}
& = na^4 \bigg[ -\frac{9H}{a^2\big[12H^2-\phi'^2\big]}{\tilde\zeta}'(\Delta_L + 4k)\tilde\zeta -\frac{3\phi'^2}{2n^2\big[12H^2 - \phi'^2\big]} {\tilde\zeta}'^2 + \notag \\
& \quad\quad\quad\;\, -\frac{3}{4a^2}\tilde\zeta(\Delta_L + 4k)\tilde\zeta - \frac{n^2}{8a^4}\frac{9}{12H^2 - \phi'^2}\tilde\zeta(\Delta_L + 4k)^2\tilde\zeta \bigg], \\
\frac{1}{\sqrt{|\gamma|}}\mathcal{L}_{\tilde\varphi\tilde\zeta}
& = na^4 \phi'\bigg[\frac{6H}{n^2\big[12H^2 - \phi'^2\big]}{\tilde\zeta}' {\tilde\varphi}' \nonumber \\
&\quad\quad\quad\;\,+ \frac{1}{2}{\tilde\varphi}'\bigg(-\frac{4}{n^2} + \frac{3}{a^2 \big[12H^2 - \phi'^2\big]}(\Delta_L + 4k)\bigg) \tilde\zeta \bigg] , 
\end{align}
and where, in analogy to cosmology, we have defined $H\equiv a'/a$.

Once we have our quadratic Lagrangians for the gauge  invariant perturbations, we can obtain the linearized equations of motion by varying the actions with respect to the corresponding fields. For the scalar sector we have
{\small\begin{align}
	0
	& = \frac{1}{a^2} \Delta_L\tilde\varphi + \frac{12}{na^4}\left(\frac{H^2 a^4}{n\big[12H^2 - \phi'^2\big]}\tilde{\varphi}'\right)' - 6\frac{1}{na^4}\left(\frac{Ha^4\phi'}{n\big[12H^2-\phi'^2\big]}\tilde{\zeta}'\right)'  + \notag \\
	& \quad - \frac{1}{2na^4}\frac{d}{d\rho}\left\{ na^4\phi' \left[-\frac{4}{n^2}+\frac{3}{a^2 \big[12H^2-\phi'^2\big]}(\Delta_L + 4k)\right]\tilde\zeta \right\},  \label{scalareom2} \\
&	\nonumber \\
	0
	& = 9\left(\frac{Hna^{2}}{\big[12H^2-\phi'^2\big]}\right)'(\Delta_L + 4k)\tilde{\zeta} -\frac{3}{2}na^{2}(\Delta_L + 4k)\tilde\zeta  \nonumber \\
	&+ 3\left(\frac{\phi'^2a^4}{n\big[12H^2-\phi'^2\big]}\tilde{\zeta}'\right)'  -\frac{9}{4}\frac{n^3 }{12H^2 - \phi'^2}(\Delta_L + 4k)^2 \tilde \zeta \nonumber\\
	&- 6\left(\frac{Ha^4\phi'}{n\big[12H^2 - \phi'^2 \big]}\tilde{\varphi}'\right)' + \frac{1}{2}n a^4 \phi' \left[-\frac{4}{n^2} + \frac{3}{a^2 \big[12H^2 - \phi'^2 \big]}(\Delta_L + 4k)\right]\tilde{\varphi}' \ . \label{scalareom1}
	\end{align}}
In analogy to cosmological perturbation theory, we can use part of the gauge freedom to set $w_i=0$ so that the vector equation of motion is simply
\begin{equation}
(\Delta_L+6k){v}^i=0  \ ,\label{vectoreom}
\end{equation}
which is solved by a vanishing vector. Meanwhile, the tensor sector is given by
\begin{equation}
\left(\frac{ a^4}{n }\left(h_{ij}^{ \ TT}\right)'\right)'+n a^2(\Delta_L+6k)h_{ij}^{TT}=0 \ . \label{tensoreom}
\end{equation}

\section{Pseudo-Conformal  universe dual}

In \cite{Hinterbichler:2014tka}, a solution of the form \eqref{domainwall} which breaks the conformal symmetries down to the $4D$ de Sitter symmetries was considered and the one-point functions were computed. This solution is sourced by a scalar with scaling dimension $\Delta=4$  with equation of motion 
\begin{equation}
\frac{d\bar{\phi}}{d\rho} = \frac{c}{a(\rho)^4} \ ,
\end{equation}
when specializing to a vanishing potential. Here, $c$ is a constant which determines the energy in the scalar field, $E_\phi \sim c^2$. 

The explicit solution for the background metric is more conveniently written in terms of a new a radial coordinate $u$ defined by $u = 1/a(\rho)^2$.   In terms of this coordinate, the metric and scalar field satisfy
\begin{align}
d\bar{s}^2 & = \frac{du^2}{4u^2(1+u+b^2 u^4)} + \frac{1}{u}\gamma_{ij}(x) dx^i  dx^j,  \label{umetric} \\
\frac{d\bar{\phi}}{du}& = -\frac{c}{2}\frac{u}{\sqrt{1+u+b^2u^4}}  \label {ufield}  \ .
\end{align}
In these equations, we have defined the dimensionless parameter 
\begin{equation}
b^2\equiv\frac{c^2\ \ell^2}{12 M^3_{\text{Pl}}} \ll 1, \label{bparameterdefee}
\end{equation}
where we have re-inserted the appropriate factors of the spatial de Sitter length $\ell$ and the higher dimensional Planck mass $M_{\text{Pl}}$ . The explicit solutions were found in \cite{Hinterbichler:2014tka} in flat Fefferman-Graham coordinates (Eq. \eqref{FG}) by working perturbatively in $b$,  and read  
\begin{align}
	\bar{g}_{t t}&=-1+\frac{b^2}{8}\left(\frac{z}{t}\right)^{8}+\mathcal{O}\left(\left(\frac{z}{t}\right)^{10}\right),\\
	\bar{g}_{11}&=1-\frac{b^2}{8}\left(\frac{z}{t}\right)^{8}+\mathcal{O}\left(\left(\frac{z}{t}\right)^{10}\right) ,\\
	\bar{\phi}&=\text { const. }-\frac{\sqrt{3}\ b}{2}\left(\frac{z}{t}\right)^{4}+\mathcal{O}\left(\left(\frac{z}{t}\right)^{6}\right).
\end{align}
We can see that the scalar field at $\mathcal{O}(b)$ back-reacts on the metric at $\mathcal{O}(b^2)$. This must be the case since, for a vanishing potential, the stress tensor depends quadratically on derivatives of $\phi$. From this solution, one can read off the one-point functions which read
\begin{equation}
\vevo_B=4\bar{\varphi}_{(0)}=-\frac{2\sqrt{3} b}{t^4} \ ,  \quad \vevt_B=2 \bar{h}_{(0) i j}=0 \ ,
\end{equation}
where $B$ indicates the background solution where all the sources have been set to zero.

For the purposes of this paper, it  is more convenient to work with the de Sitter slicing in the $u$-coordinates of Eq.\eqref{umetric}  instead of the flat Fefferman-Graham coordinates. In this case, the asymptotic expansion of the fields is
\begin{align}
g_{ij}(x, u)&=\hat{g}_{(0)\,  ij}+u  \, \hat{g}_{(1)\,  ij}+u^{2}\left(\hat{h}_{(0)\, ij}+\hat{g}_{(2)\, ij} \log{u}+\cdots\right)+\cdots \ ,  \\
\phi(x, u)&=\hat{\phi}_{(0)}+u  \, \hat{\phi}_{(1)}+u^{2} \left(\hat{\varphi}_{(0)}+\hat{\phi}_{1(2)} \log{u}\right)+\cdots \ .
\end{align}
Near the boundary we have $u\sim z^2/t^2+z^4/t^4$ which can be used in the results above to relate the asymptotic expansions in the $u$ and  $z$-coordinates.  The one-point functions in this slicing can be obtained by using Eq.\eqref{Weylt} with $\Omega=1/t^4$, yielding
\begin{equation}
\vevo_\gamma=4 t^4 \varphi_{(0)}=4\left(\hat{\phi}_{(1)}+\hat{\varphi}_{(0)}\right)\ ,  \quad \vevt_\gamma=2 t^4 h_{(0) \, i j}=2  \left(\hat{g}_{(1) \, i j}+\hat{h}_{(0) \, i j}\right) \ , \label{onepoint}
\end{equation} 
and the background values are
\begin{equation}
\left(\vevo_\gamma\right)_B=-2 \sqrt{3} b\ ,  \quad \left(\vevt_\gamma\right)_B=0 \ . \label{onepointb}
\end{equation}
In the following, we focus on finding the solutions to the linearized perturbed equations and explicit expressions for the two-point correlators.

\subsection{Perturbations and two-point correlators}
The two-point functions of a CFT where the conformal symmetries are spontaneously broken due to a VEV down to $\mathfrak{so}(1,4)$ are obtained by solving the linearized equations of motion for the perturbations around the solution in Eqs. \eqref{umetric} and \eqref{ufield}. These are given by
\begin{align}
\left\langle\op(x)\op(x^{\prime})\right\rangle_{\eta}&=-\frac{1}{t^4}\frac{1}{t^{\prime  4}}\sum_{\lambda} Y_{\lambda}(x) Y_{\lambda}\left(x^{\prime}\right) \  \frac{\delta \vevo_\gamma}{\delta  \varphi_{(0)}}(\lambda) \ ,\\
\left\langle\stresst^{TT}(x)T_{kl}^{TT}(x^{\prime})\right\rangle_{\eta}&=2\frac{1}{t^4}\frac{1}{t^{\prime  4}}\sum_{\lambda} Y_{\lambda}(x) Y_{\lambda}\left(x^{\prime}\right)  \frac{\delta \left\langle T_{ij} \right\rangle_\gamma}{\ \delta  {h^{TT}_{(0)}}^{ \ kl}}(\lambda)\ ,.
\end{align}
where the VEVs are given by Eq.\eqref{onepoint}.

In Appendix \ref{embeddingappendixe} we determine the extent to which the breaking pattern \eqref{symbreakpatternee} determines the forms of these correlators in the pseudo-conformal universe.  Note that any function of the form $\sum_\lambda Y_{\lambda}(x) Y_{\lambda}\left(x^{\prime}\right) F(\lambda)$ is invariant under the $dS_4$ invariant distance between the points $x$ and $x'$.  On flat space this becomes precisely the cross ratio \eqref{crossrationeqe}.  Thus the form of the scalar-scalar correlator will have the correct form \eqref{finalsscformee} for a scalar of dimension $\Delta=4$.

We now find the solutions to the linearized equations  of motion for the perturbations around the domain wall background reviewed above. We start by taking the following ansatz for the perturbations
\begin{align}
\tilde{\varphi}(x,u)&=\sum_{\lambda} \tilde{\varphi}_\lambda(u) Y_{\lambda}(x) \ , \\
\tilde{\zeta}(x,u)&=\sum_{\lambda} \tilde{\zeta}_\lambda(u) Y_{\lambda}(x)  \ , \\
\tilde{h}_{ij}^{TT}(x,u)&=e_{ij}\sum_{\lambda} \tilde{f}_\lambda(u) Y_{\lambda}(x)  \  ,
\end{align}
which is just an expansion over the basis $Y_\lambda$. Here, $e_{ij}$ is the polarization tensor of the graviton. Note that we only Fourier transform the CFT coordinates and leave the radial coordinate untouched. The equations of motion for the perturbations in the $u$-coordinates can be obtained from Eqs.\eqref{scalareom2},\eqref{scalareom1}, 
and \eqref{tensoreom} by setting 
\begin{equation}
k=1, \quad \phi' = -\frac{c}{2}\frac{u}{\sqrt{1+u+b^2u^4}}, \quad  n^2 =\frac{1}{4u^2(1+u+b^2u^4)}, \quad  a^2 = \frac{1}{u}, \quad  H = -\frac{1}{2u} \ .
\end{equation}
For the coupled scalar sector they read
\begin{align}
&\frac {\lambda  \varphi (u)} {2 u^2 \sqrt {G (u)}} + 
2\partial_u\Bigg (\frac {G (u)^{3/2}\varphi' (u)} {u (b^2 - 
	G (u))}\Bigg) -2\partial_u\Bigg(\frac {G (u)\zeta' (u)} {u \
	(b^2 - G (u))}\Bigg) \nonumber \\
&+4\sqrt {3}b\zeta' (u)+\sqrt {3 b}\partial_u\Bigg(\zeta (u)\left (\frac{-\lambda+ 1}{4u^3 (b^2 - G (u))} + 4 \right)\Bigg)=0 \\
&(-\lambda +4) \zeta (u) \left(9\partial_u \left(\frac{\sqrt{G(u)}}{12 u^3 (G(u)-b^2)}\right)+\frac{3}{4 u^2 \sqrt{G(u)}}\right)+\frac{3 (-\lambda +4)^2 \zeta (u)}{32 u^5 \sqrt{G(u)} (b^2-G(u))}  \nonumber \\
&-\frac{2 \sqrt {3 }b G(u) \varphi ''(u)}{u (b^2-G(u))} -4 \sqrt {3 }b \varphi '(u) \left(\frac{-\lambda +4}{4 u^3
	(b^2-G(u))}+4\right)+6b \partial_u\left(\frac{\sqrt{G(u)} \zeta '(u)}{u (b^2-G(u))}\right) \nonumber \\
&2 \sqrt {3 }b\partial_u\left(\frac{G(u)}{u (b^2-G(u))}\right)=0\,,
\end{align}
where $G(u)=1+u+b^2 u^4$ . Similarly, for the tensor sector we have 
\begin{align}
4 \left(1+u+b^2 u^4\right) \left(u
\tilde{f}_\lambda''(u)-\tilde{f}_\lambda'(u)\right)+ (2+\lambda )\tilde{f}_\lambda(u) &=0 \ .
\end{align}
The vector equation of motion is easily solved for an arbitrary $\lambda$ by a vanishing $v^i$, which is similar to the situation in cosmological perturbation theory. 
Therefore, in the following, we focus on solving the tensor and scalar sectors and can ignore the vectors.

We will solve for $\tilde{\varphi}_{\lambda}(u), \tilde{\zeta}_{\lambda}(u),$ and $f_{\lambda}(u)$ by finding a solution as a series expansion in the small parameter $b$. The ansatz for expansions of the fields are
\begin{align}
\tilde{\varphi}&= \tilde{\varphi}^1 b + \tilde{\varphi}^3 b^3 + \cdots \ , \\
\tilde{\zeta}&= \tilde{\zeta}^2 b^2+ \tilde{\zeta}^4 b^4+\cdots \ , \\
f&=f^2 b^2+ f^4 b^4+\cdots \ ,
\end{align}
which is motivated by the fact that a scalar field of order $b$ backreacts on the metric at order $b^2$, which in turn backreacts on the scalar at  order $b^3$. This trend is followed to all higher orders in $b$. By solving the equations of motion order by order in $b$, we find that the leading order solutions are given by 
{\footnotesize
\begin{align}
\tilde{\varphi}=& - b\  \frac{\pi}{8}\  \varphi^1_{(0)}   l^2  \lambda \sqrt{4+ l^2  \lambda} \left(u+1\right)
 \csc \left(\frac{1}{2} \pi 
\sqrt{4+ l^2  \lambda}\right) \, _2F_1\left(-\frac{1}{2} \sqrt{4+ l^2  \lambda
},\frac{1}{2} \sqrt{4+ l^2  \lambda };2;u+1\right) \ , \\
 \tilde{\zeta}=& \ b^2\frac{32 z^4 \sqrt{u+1} \left(\left(-t^8 \left( l^2  \lambda+4\right)+2 t^6 z^2+2 t^4 z^4-16
 	t^2 z^6-16 z^8\right) \tilde{\varphi}'\left(u\right)-2 t^2 z^4 \left(t^2+z^2\right)
 \tilde{\varphi}''\left(u\right)\right)}{\sqrt{3} t^4 \left( l^2  \lambda+4\right)
 	\left(-t^8 \left( l^2  \lambda+4\right)+24 t^6 z^2+24 t^4 z^4-8 t^2 z^6-8 z^8\right)}  \ , \\
  f=& b^2  f^2_{(0)}   \frac{\pi}{8}  (u+1) \ (l^2\lambda+2) \sqrt{\! -l^2\lambda+2}  \ {}_2F_1\left(\!\!-\frac{1}{2} \sqrt{\! -l^2\lambda+2}, \ \frac{1}{2} \sqrt{\!-l^2\lambda+2};2;u+1\right) \csc \left(\frac{1}{2} \pi  \sqrt{\!-l^2\lambda+2}\right)  ,
\end{align}
} 
where we have required that the solutions remain finite inside the bulk. The boundary conditions imposed on these solutions were found by looking at the gauge invariant combination
\begin{equation}
\tilde{\varphi}-\frac{\phi'}{2 H}\tilde{\zeta}={\varphi}-\frac{\phi'}{2 H}{\zeta} \ .
\end{equation}
Note that terms involving $\zeta_0$ will first appear at order $b^3$. More details can be found in Appendix \ref{ap1}. Similarly, for the metric perturbation we have $f(z=0)=f_0$.  The solutions above are complex -- we take the real part to obtain a real solution.\footnote{The imaginary part will have no dependence on the source, and thus is not considered here since it makes no contribution to the correlators.}

In order to compute the two point functions we can look at the asymptotic expansion of the perturbations. More specifically, we seek the linear terms in the sources of the $u$ and $u^2$ coefficients. Here, we will only show the final results and leave the details of this computation for  Appendix \ref{ap1}. The two-point functions at leading order in $b$ read 
\begin{align}
\left\langle\op(x)\op(x^{\prime})\right\rangle_{\eta}=&- b\frac{1}{t^4}\frac{1}{t^{\prime  4}}\sum_{\lambda} Y_{\lambda}(x) Y_{\lambda}\left(x^{\prime}\right) \frac{ \left( l^2  \lambda+4\right)  }{4} l^2\lambda\operatorname{Re}\left(   H_{1-\frac{1}{2} \sqrt{4+l^2  \lambda }}+   H_{1+\frac{1}{2} \sqrt{4+l^2  \lambda }}\right)\! \nonumber \\
&+ \text{local terms} \ ,\\
\left\langle\stresst^{TT}(x)T_{kl}^{TT}(x^{\prime})\right\rangle_{\eta}=&b^2\frac{1}{t^4}\frac{1}{t^{\prime  4}}\sum_{\lambda} \Pi_{i j k l}^{T T} \ Y_{\lambda}(x) Y_{\lambda}\left(x^{\prime}\right)   \frac{(4-l^4 \lambda^2)}{16} \operatorname{Re}\left(H_{1-\frac{1}{2} \sqrt{18-l^2\lambda}}+H_{1+\frac{1}{2} \sqrt{18-l^2\lambda}} \right) \nonumber \\
& + \text{local terms}  \ ,
\end{align}
where $H_n$ is the $n$-th Harmonic number defined as $H_{n}=\sum_{i=1}^{n} 1$, and can be expressed analytically as 
\begin{equation}
H_n=\gamma+\Psi(n+1)=\gamma+\frac{\Gamma'(n+1)}{\Gamma(n+1)} \ ,
\end{equation} 
where $\gamma$ is the Euler constant, $\Psi(n)$ is the digamma function, and $\Gamma(n)$ is the gamma function.  We  have also defined the transverse-traceless projector $\Pi_{i j k l}^{T T}$ as 
\begin{align}
\Pi_{i j k l}^{T T} &\equiv\frac{\delta h_{(0) i j}^{TT}}{\delta h_{(0)}^{TT \ k l}}=\frac{1}{2}\left(\pi_{i k} \pi_{j l}+\pi_{i l} \pi_{j k}\right)-\frac{1}{3} \pi_{i j} \pi_{k l} \, 
\end{align}
in terms of the projection operator $\pi_{i j}=\delta_{i j}-\nabla_{i} \nabla_{j} / \nabla^{2}$. 

\section{Conclusions}

We have shown how to holographically compute two-point correlators for a CFT in which the conformal symmetries are spontaneously broken down to a de Sitter $\mathfrak{so}(4,1)$ algebra by a space-like defect. This models the pseudo-conformal universe, an alternative to inflation.  This is one of three ways of breaking conformal symmetry down to a subgroup corresponding to isometries of a maximally symmetric space of the same dimension as the CFT.   The others are breaking to the Poincar\'e group $\mathfrak{iso}(3,1)$ or to the anti-de Sitter group $\mathfrak{so}(3,2)$.  The case of symmetry breaking down to Poincar\'e occurs when a scalar gets a constant VEV, and has been widely analyzed in the holographic context, for example in \cite{Bianchi:2001de,DeWolfe:2000xi}. 
The breaking down to anti-de Sitter corresponds to a CFT with a codimension-1 timelike defect.  Some examples analyzing this breaking can be found in \cite{Gutperle:2017nwo} (see also \cite{Chiodaroli:2016jod} for a computation of scalar two-point functions for conformal interfaces in the D1/D5 CFT.)

The advantage of this holographic approach, as opposed to the direct approach of \cite{Hinterbichler:2011qk} is that it can model a strongly coupled scenario.  Generally, inflation and other early universe scenarios are taken to be weakly coupled so that they remain calculable.   The pseudo-conformal scenario, by contrast, is built on a CFT, which can be strongly coupled, and thus accessible through AdS/CFT.   Another interesting direction would be to adapt the conformal bootstrap \cite{Rattazzi:2008pe,Poland:2018epd,Simmons-Duffin:2016gjk} to the pseudo-conformal case, in order to directly explore strongly coupled possibilities.  The defect conformal bootstrap has already seen much development
\cite{McAvity:1993ue,McAvity:1995zd,Cardy:2004hm,Liendo:2012hy,Gliozzi:2015qsa,Billo:2016cpy,deLeeuw:2017dkd,Bissi:2018mcq,Behan:2020nsf}, and adapting this to the pseudo-conformal case would presumably put constraints on the possible CFTs and operators that could realize the scenario.

We do not necessarily expect the scalar two-point function we computed to correspond directly to the power spectrum that is observed in the CMB.  As discussed in \cite{Hinterbichler:2011qk}, the observed scale-invariant two-point function should instead come from some $\Delta=0$ spectator field, whose perturbations must then be imprinted on the CMB \cite{Wang:2012bq}.  It would be interesting to extend the calculation to include such spectators and create a more realistic model, perhaps in a setup similar to \cite{Hinterbichler:2012fr,Hinterbichler:2012yn}.

{\bf Acknowledgements:} KH acknowledges support from DOE grant DE- SC0019143 and Simons Foundation Award Number 658908. The work of MT is supported in part by US Department of Energy (HEP) Award DE- SC0013528, and by the Simons Foundation Origins of the Universe Initiative, grant number 658904.

\appendix
\section{Series expansions and scalar two-point function} \label{ap1}
In order to find a solution to the linearized equations of motions for small fluctuations about the domain wall background, we consider series expansions in the small parameter $b$ defined in \eqref{bparameterdefee}, so that the fields read
\begin{align}
\tilde{\varphi}&= \tilde{\varphi}^1 b + \tilde{\varphi}^3 b^3 + \cdots \ , \\
\tilde{\zeta}&= \tilde{\zeta}^2 b^2+ \tilde{\zeta}^4 b^4+\cdots \ , \\
{\varphi}&= {\varphi}^1 b + {\varphi}^3 b^3 + \cdots \ , \\
{\zeta}&= {\zeta}^2 b^2+ {\zeta}^4 b^4+\cdots \ , \\
f&=f^2 b^2+ f^4 b^4+\cdots \ .
\end{align}
The leading order power in $b$ is justified by solving the equations of motion and by the behavior of the background solution. Besides the expansion in $b$, we also perform an expansion in $u$ to find the boundary conditions and two-point functions. We will specifically look at the near-boundary expansion of the gauge invariant combination
\begin{equation}
\tilde{\varphi}-\frac{\phi'}{2 H}\tilde{\zeta}={\varphi}-\frac{\phi'}{2 H}{\zeta} \ .
\end{equation}
\paragraph{Order $u^0$}
This order sets the  boundary conditions for the gauge invariant fields. To find this, we expand 
\begin{equation}
\frac{\phi'}{2 H}=b \sqrt{\frac{3}{1+u+b^2  u^4}}
\end{equation}
around $u=0$. It is easy to see that at leading order in $b$ we have
\begin{equation}
\tilde{\varphi}^1_{(0)}=\varphi^1_{(0)}  \ ,
\end{equation}
where the index $(n)$ labels the coefficient of the order $u^n$ term in the series expansion, and the upper index $m$ labels the coefficient of the order $b^m$ term. Terms containing $\zeta_0$ will appear at order $b^3$ and lead to
\begin{equation}
\tilde{\varphi}^3_{(0)}=\varphi^3_{(0)}-\sqrt{3} b \zeta^2_{(0)} \ ,
\end{equation}
where we have used the fact that in our perturbative solution $\tilde{\zeta}^2_{(0)}=0 $. Similarly, at order $b^n$ we have
\begin{equation}
\tilde{\varphi}^n_{(0)}={\varphi}^n_{(0)}-\sqrt{3}\left(\zeta^{n-1}_{(0)}-\tilde{\zeta}^{n-1}_{(0)}\right) \ . \label{order0}
\end{equation}
In order to make the dependence on $b$ explicit, the derivatives with respect to the sources will now be written as
\begin{equation}
\frac{\delta}{\delta \varphi_{(0)}}=\frac{1}{b}\frac{\delta}{\delta \varphi^1_{(0)}}+\frac{1}{b^3}\frac{\delta}{\delta \varphi^3_{(0)}}+\cdots
\end{equation}
\paragraph{Order $u$}
At this order we have 
\begin{equation}
\sqrt{3}b{\zeta}_{(1)}={\varphi}_{(1)}-\tilde{\varphi}_{(1)}+\sqrt{3}b\tilde{\zeta}_{(1)}+\frac{1}{2 }\sqrt{3}b\left(\zeta_{(0)}-\tilde{\zeta}_{(0)}\right) \ . \label{z2}
\end{equation}
From this we see that at leading order in $b$
\begin{align}
\sqrt{3}b\frac{\delta \zeta_{ (1)}}{\delta \varphi_{(0)}}&=\frac{\delta( {\varphi}^3_{ (1)}- \tilde{\varphi}^3_{ (1)}))}{\delta \varphi^3_{(0)}}+\frac{\delta( {\varphi}^5_{ (1)}- \tilde{\varphi}^5_{ (1)}))}{\delta \varphi^5_{(0)}}+\cdots \,,  \\
\sqrt{3}b\frac{\delta \zeta_{ (1)}}{\delta \zeta_{(0)}}&=\frac{\delta( {\varphi}^3_{ (1)}- \tilde{\varphi}^3_{ (1)}))}{\delta \zeta^2_{(0)}}+\frac{\delta( {\varphi}^5_{ (1)}- \tilde{\varphi}^5_{ (1)}))}{\delta \zeta^4_{(0)}}+\cdots+\frac{1}{2 }\sqrt{3}b \ .
\end{align}
Note that any contribution from $\tilde{\zeta}$ is higher order since we know from solving the equations of motion order by order  that $\tilde{\zeta}^n=\tilde{\zeta}^n(\varphi^{n-1}_{(0)},\varphi^{n-3}_{(0)},\cdots,\zeta^{n-2}_{(0)},\cdots)$. Also, at leading order in $b$ the relation  in Eq.\eqref{z2} becomes ${\varphi}^1_{(1)}-\tilde{\varphi}^1_{(1)}=0$.
\paragraph{Order $u^2$}
This order determines the two-point functions and reads
\begin{equation}
{\varphi}_{(2)}=\tilde{\varphi}^1_{(2)}b+\tilde{\varphi}^3_{(2)}b^3+\cdots+\sqrt{3} b\left({\zeta}_{(2)}-\tilde{\zeta}_{ (2)}\right)-\frac{\sqrt{3} b }{2}\left({\zeta}_{ (1)}-\tilde{\zeta}_{ (1)}\right)+\frac{3}{8}\sqrt{3}b(\zeta_{(0)} -\tilde{\zeta}_{(0)})\ .
\end{equation}
At leading order in $b$ we find
\begin{align}
\frac{\delta \varphi_{ (2)}}{\delta \varphi_{(0)}}&=\frac{\delta \tilde{\varphi}^1_{ (2)}}{\delta \varphi^1_{(0)}}+\frac{\delta \tilde{\varphi}^3_{ (2)}}{\delta \varphi^3_{(0)}}+\cdots-\frac{1}{2}\left(\frac{\delta( {\varphi}^3_{ (1)}- \tilde{\varphi}^3_{ (1)}))}{\delta \varphi^3_{(0)}}+\frac{\delta( {\varphi}^5_{ (1)}- \tilde{\varphi}^5_{ (1)}))}{\delta \varphi^5_{(0)}}+\cdots\right) \ , \label{dphi4dphi}\\
\frac{\delta \varphi_{ (2)}}{\delta \zeta_{(0)}}&=\left[\frac{\delta \tilde{\varphi}^3_{ (2)}}{\delta \zeta^2_{(0)}}+\frac{\delta \tilde{\varphi}^5_{ (2)}}{\delta \zeta^4_{(0)}}+\cdots-\frac{1}{2}\left(\frac{\delta( {\varphi}^3_{ (1)}- \tilde{\varphi}^3_{ (1)}))}{\delta \zeta^2_{(0)}}+\frac{\delta( {\varphi}^5_{ (1)}- \tilde{\varphi}^5_{ (1)}))}{\delta \zeta^4_{(0)}}+\cdots\right)+\frac{1}{8}\sqrt{3}\right]b \ , \label{dphi4dzeta} 
\end{align}
where we have used the expansions at lower orders in $u$ to simplify these expressions. We have also used the fact that $\zeta_{(2)}\sim\left\langle T_i^i\right\rangle$ and the Ward Identity, Eq. \eqref{tracet}. We have omitted any terms that are not linear in the sources, since these would not contribute to the 2-point function. In principle, such terms arise from the conformal anomaly contribution in Eq. \eqref{tracet}. \\

We proceed to simplify the scalar two-point function result by using the Ward Identities. Combining Eq.\eqref{fromwi} and Eq.\eqref{dphi4dzeta} we find  
\begin{equation}
\frac{\delta \tilde{\varphi}^3_{ (2)}}{\delta \zeta^2_{(0)}}+\frac{\delta \tilde{\varphi}^5_{ (2)}}{\delta \zeta^4_{(0)}}+\cdots+\frac{1}{2}\left(\frac{\delta(3 {\varphi}^3_{ (1)}+ \tilde{\varphi}^3_{ (1)}))}{\delta \zeta^2_{(0)}}+\frac{\delta(3 {\varphi}^5_{ (1)}+ \tilde{\varphi}^5_{ (1)}))}{\delta \zeta^4_{(0)}}+\cdots\right)=\frac{7}{8 }\sqrt{3} \ .
\end{equation}
Now, we can use the fact that the sources will appear only in the combination on the right hand side of Eq.\eqref{order0} to write 
\begin{equation}
\frac{\delta \tilde{\varphi}^3_{ (2)}}{\delta  \varphi^3_{(0)}}+\cdots+\frac{1}{2}\left(\frac{\delta( 3{\varphi}^3_{ (1)}+ \tilde{\varphi}^3_{ (1)}))}{\delta \varphi^3_{(0)}}+\frac{\delta(3 {\varphi}^5_{ (1)}+ \tilde{\varphi}^5_{ (1)}))}{\delta \varphi^5_{(0)}}+\cdots\right)=\frac{7}{8 } \ .
\end{equation}
This allows us to write the two-point function at leading order in $b $ as 
\begin{equation}
\left\langle\op(\lambda)\op(\lambda)\right\rangle=-4\frac{\delta \tilde{\varphi}^1_{ (2)}}{\delta \varphi^1_{(0)}}-\frac{7}{2} \ .
\end{equation}
Notice that this final expression only requires us to find the leading order solution of the perturbed equations, unlike in Eq.\eqref{dphi4dphi}.

\section{Pseudo-conformal correlators from embedding space\label{embeddingappendixe}}

Correlation functions of the type we are interested in are heavily constrained by the symmetry breaking pattern \eqref{symbreakpatternee}.
Here we present a method to find the form of correlators which satisfy these constraints, directly from symmetry considerations.  This relies on the embedding space approach for finding conformally covariant correlators \cite{Dirac:1936fq,Weinberg:2010fx,Costa:2011mg}, extended to allow for spontaneous symmetry breaking \cite{Fubini:1976jm,Liendo:2012hy,Billo:2016cpy,Sarkar:2018eyy}.  (for other earlier work on CFT correlators near a boundary see \cite{Cardy:1984bb,McAvity:1993ue,McAvity:1995zd,Lauria:2017wav}.)

We consider an ambient $6$ dimensional Lorentzian space, with cartesian coordinates
\be X^A=\{ X^{-1},X^{0},X^\mu\}\, ,\ee 
and metric 
\be \eta_{AB}=diag(-1,1,\eta_{\mu\nu}),\label{ambeintmetric}\ee
so that $X^{-1}$ is a ``time" dimension, and $\mu$ labels the $4$ dimensions of the CFT.  

Now consider the lightcone, centered at the origin,
\be \eta_{AB}X^AX^B=-(X^{-1})^2+(X^0)^2+\eta_{\mu\nu}X^\mu X^\nu =0.\label{coneconditione}\ee
The space of rays of the lightcone, the equivalence classes of points on the lightcone under the identification
\be X^A\sim \lambda X^A,\ \ \ \lambda \in {\mathbb R},\ee
is the {conformal sphere}, the space on which the CFT lives.
Flat space is given by choosing the ray representative of each point of the conformal sphere which is given by the intersection with the plane 
\be X^{-1}+X^0=1.\label{flatsliceplanee}\ee
(This misses one point on the conformal sphere, the ray at the south pole, $X^{-1}=-X^0$, $X^\mu=0$.  This will be the point at infinity.)  The surface on which the CFT lives is thus embedded as
\be X^{-1}(x)={1\over 2}(1+x^2),\ \ \ X^{0}(x)={1\over 2}(1-x^2),\ \ \ X^\mu(x)=x^\mu,\label{embedding1e}\ee
and the induced metric is the usual flat metric,
\be \partial_\mu X^A\partial_\nu X^B\eta_{AB}=\eta_{\mu\nu}.\ee
Lorentz transformations in the ambient space become precisely the conformal transformations in the embedded flat space.

It is convenient to use light cone coordinates in embedding space:
\be X^{\pm}=X^{-1}\pm X^0,\ \ X^{-1}={1\over {2}}\left(X^++X^-\right),\ \ \ X^{0}={1\over {2}}\left(X^+-X^-\right).\ee
The embedding space metric is now
\be ds^2=-dX^+dX^-+\eta_{\mu\nu}dx^\mu dX^\nu, \ee
the light cone condition \eqref{coneconditione} becomes
\be - X^+X^-+\eta_{\mu\nu}X^\mu X^\nu=0,\ee
and the flat slice condition \eqref{flatsliceplanee} becomes
\be X^+=1.\ee
The embedding \eqref{embedding1e} becomes $X^+(x)=1,\ X^-(x)=x^2,\ X^\mu(x)=x^\mu,$
i.e. 
\be X^A(x)=(1,x^2,x^\mu).\ee

There is a bijective map between symmetric tensor fields $t_{\mu_1\cdots \mu_s}(x)$ of rank $s$ in the CFT and symmetric tensor fields $T_{A_1\cdots A_s}(X)$ of rank $s$ in the ambient space that satisfies the following conditions:
\begin{align} 
& {\rm defined\ on\ the\ null\ cone\ } X^2=0 , \\
& {\rm homogeneity:} \quad\left( X^A \partial_A+\Delta \right)T_{A_1\cdots A_s}=0 \Leftrightarrow  T_{A_1\cdots A_s}(\lambda X)=\lambda^{-\Delta} T_{A_1\cdots A_s}(X), \label{scalingcondine}\\
&{\rm tangentiality:}  \quad~X^{A_1}  T_{A_1\cdots A_s}( X)=0\, ,\label{tangentialityconde}
\end{align}
where $\Delta$ is some fixed real number, the homogeneity degree.  \eqref{scalingcondine} and \eqref{tangentialityconde} serve to define the tensor everywhere on the cone, given that they are defined on the flat section.
The fact that $T_{A_1\cdots A_s}(X)$ is defined on the null cone means that its value off the cone is unimportant, which translates to requiring the gauge symmetry-like identification 
\be T_{A_1\cdots A_s}(X)\sim T_{A_1\cdots A_s}(X)+X_{(A_1}U_{A_2\cdots A_s)}+X^2 V_{A_1\cdots A_s},\label{gaugeidente}\ee
for any symmetric $U_{A_1\cdots A_{s-1}}$, $V_{A_1\cdots A_s}$.  

The relation between the tensors is given by the pullback,
\be t_{\mu_1\cdots \mu_s}(x)=e^{A_1}_{\mu_1}\cdots e^{A_s}_{\mu_s}T_{A_1\cdots A_s}(X(x))\,.\label{projectionmape}\ee
where the tangent vectors to the embedding surface are
\be e^A_\mu\equiv{\partial X^A\over \partial x^\mu}=(0,2x_\mu,\delta^\nu_{\ \mu}).\ee
Under conformal transformations, i.e. Lorentz rotations in the ambient space, $t_{\mu_1\cdots \mu_s}(x)$ transforms like a conformal primary of dimension $\Delta$.

It is generally convenient to package symmetric traceless tensor operators using polarizations $z^\mu$,
\be t(x,z)=t_{\mu_1\cdots\mu_s}z^{\mu_1}\cdots z^{\mu_s}.\ee
The polarizations satisfy $z^2=0$, which ensures tracelessness of the operator.  Ambient space operators are contractions with $6D$ polarizations $Z^A$,
\be T(X,Z)=T_{A_1\cdots A_s}Z^{A_1}\cdots Z^{Z_s}.\ee
The relation between the two is
\be Z^A=(0,2x\cdot z,z^\mu).\ee
This satisfies the conditions $Z^2=0$, $Z\cdot X=0$, which ensure tracelessness and the transversality condition.   

Given two points $1$ and $2$, the following relations are useful when reducing from ambient space to CFT space,
\be X_1\cdot X_2=-{1\over 2}(x_1-x_2)^2,  \ \ \ Z_1\cdot Z_2=z_1\cdot z_2,\ \ \ X_1\cdot Z_2=(x_1-x_2)\cdot z_2\,.\label{projectrelationsube}\ee

In an unbroken CFT, the correlators must be transverse functions of two embedding space fields with the correct scaling.  This fixes all the one-point functions to vanish (there is no non-vanishing transverse structure that can be made from a single point), and fixes the 2-pt. functions to be diagonal
\be \la T(X_1,Z_1)T(X_2,Z_2)\ra\sim {H^s\over (X_1\cdot X_2)^{\Delta}}.\ee
Here $s$ is the spin of the two operators and $\Delta$ is their scaling dimension (the correlator must vanish if the spins or scaling dimensions of the two operators are different).
\be H\equiv  { Z_1\cdot Z_2}-{X_1\cdot Z_2\, X_2\cdot Z_1\over X_1\cdot X_2\,}\ee
is the unique transverse, scale-invariant structure of two points. 
Upon using \eqref{projectrelationsube} the correlator projects to 
\be \la t(x, z_1)t(0, z_2)\ra=\frac{\left[ z_1\cdot z_2 \, x^2-2(x\cdot z_1)\,  (x\cdot z_2)\right]^s}{x^{2(\Delta+s)}}.\label{spins2ptzse}\ee

\begin{figure}[h!]
\begin{center}
\epsfig{file=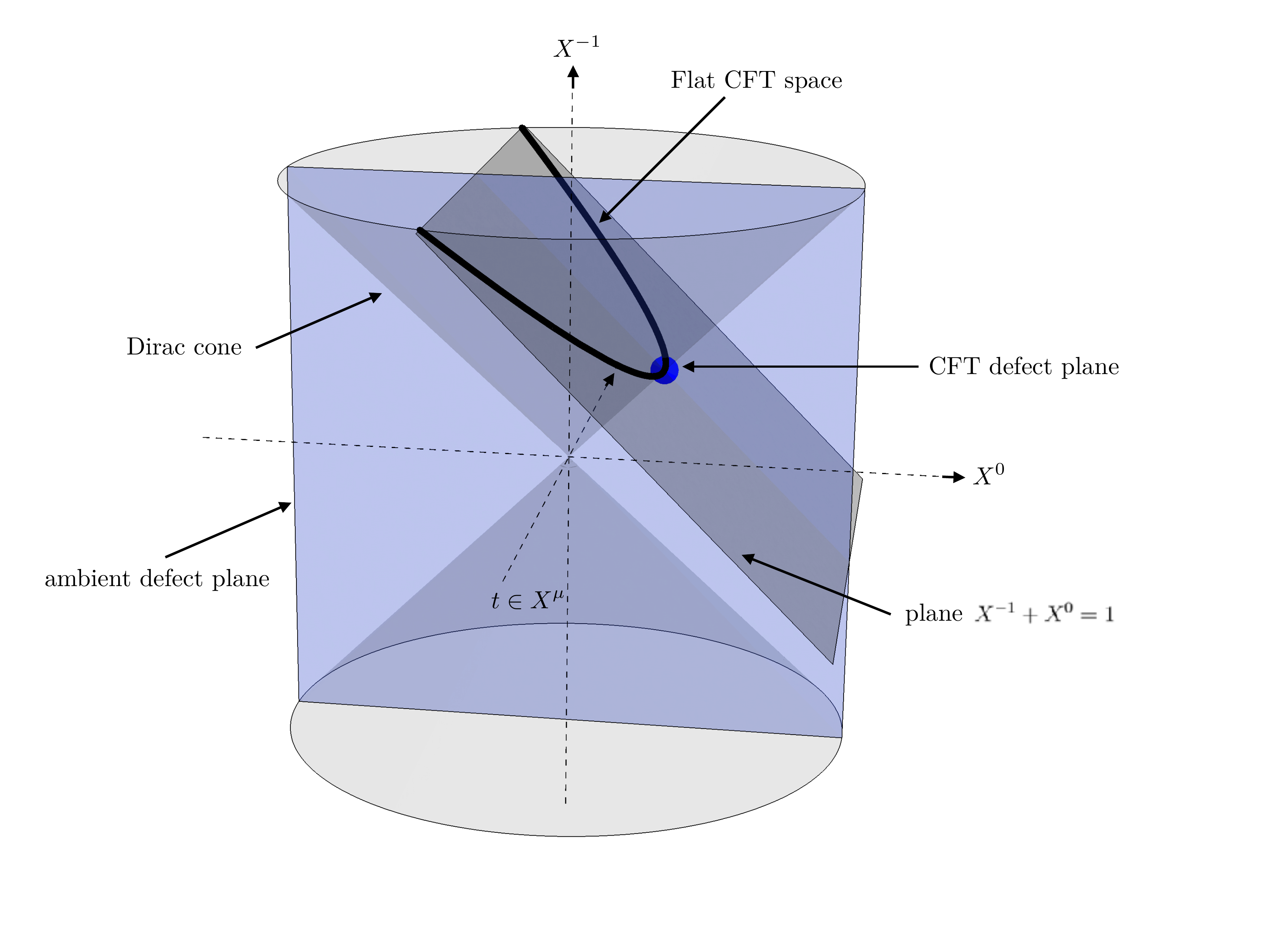,height=5in,width=6.0in}
\caption{Embedding space for the pseudo-conformal correlators.  The vector field $V^A$ is normal to the ambient defect plane $X^0=0$, whose intersection with the CFT surface is the defect plane at $t=0$.}
\label{plot1}
\end{center}
\end{figure}

In the presence of a space-like defect (the $t=0$ boundary of our pseudo-conformal universe) there is a preferred direction in the CFT --- the direction normal to the defect.  This can be represented as a unit-normalized vector field $v^\mu=(1,0,0,0)$ in the CFT.  We can extend this to an embedding space vector field
\be V^A=(0,0,1,0,0,0),\ee
which can now be used as an ingredient in constructing the invariants for correlators, see figure \ref{plot1}.  On the hypersurface we have
\be X\cdot V=x\cdot v=(-t),\ \ \ Z\cdot V=z\cdot v=-z^t\label{tfromdote}\ee

\subsubsection{One-point functions}

In an unbroken vacuum the one-point functions vanish because no invariants can be constructed from a single $X$ or $Z$.  However, in the broken vacuum we can now use $V$, and 
there is a single scalar invariant that can be constructed from a single $X^A$ and $V^A$, namely  $X\cdot V.$  Considering a scalar one-point function $\la \Phi(X)\ra\sim (X\cdot V)^p,$
for some power $p$, the scaling requirement \eqref{scalingcondine}, $\Phi(\lambda X)=\lambda^{-\Delta}\Phi(X)$, tells us that $ \Delta=-p$. Thus, the scalar 1-point function is
\be \la \Phi(X)\ra\sim {1\over (X\cdot V)^\Delta}. \ee
Restricting to the flat hypersurface using \eqref{tfromdote}, this is
\be { \la\phi(x)\ra \sim {1\over (-t)^{\Delta}},}\ee
the correct form \eqref{vevforminte} for the pseudo-conformal universe symmetry breaking VEV.  

Higher spin operators ($s>0$) are unable to develop VEVs because the numerator would need to have a factor $(V\cdot Z)^s$ which is not transverse, and so the one-point functions vanish,
\be \la \phi_{\mu_1\cdots \mu_s}(x)\ra=0,\ \ \ s\geq 1.\ee

\subsubsection{Scalar-scalar 2-point functions}

There are three scalar invariants that can be made from the two points, 
\be X_1\cdot X_2,\ \ \ X_1\cdot V,\ \ \ X_2\cdot V.\ee
The two point function should be a function of these, $\la \Phi_1(X_1)\Phi_2(X_2)\ra\sim (X_1\cdot X_2)^{p_1} (X_1\cdot V)^{p_2} (X_2\cdot V)^{p_3},$
and the scaling requirements $\Phi_1(\lambda X_1)=\lambda^{-\Delta_1}\Phi_1(X_1)$, $\Phi_2(\lambda X_2)=\lambda^{-\Delta_2}\Phi_2(X_2)$ 
give $ p_1+p_2=-\Delta_1,\ \ \ p_1+p_3=-\Delta_2$. This is a rank 2 system for the 3 $p$'s, so there is 1 undetermined invariant (cross-ratio) which is not fixed.
We may choose the particular solution $p_2=-\Delta_1,\ \ p_3=-\Delta_2,$ 
and the  homogenous solution, $p_i=(1,-1,-1),$
corresponding to the cross ratio 
\be \xi\equiv -{1\over 2}{X_1\cdot X_2\over (X_1\cdot V)(X_2\cdot V)}={1\over 4}{(x_1-x_2)^2\over (- t_1)(- t_2)}.\label{crossrationeqe}\ee
The correlation function may contain an arbitrary function $F$ of this cross ratio, and thus takes the form 
\be \la \Phi_1(X_1)\Phi_2(X_2)\ra\sim { F(\xi)\over (X_1\cdot V)^{\Delta_1} (X_2\cdot V)^{\Delta_2}},\ee
which, after reducing to the flat hypersurface, gives 
\be { \la \phi_1(X_1)\phi_2(X_2)\ra\sim {F(\xi)\over (-t_1)^{\Delta_1}(-t_2)^{\Delta_2}} }.\label{finalsscformee}\ee
This matches the form computed in special cases in \cite{Hinterbichler:2011qk,Hinterbichler:2012mv}.

Note that the conformal weights of the two scalars may be different.  This is unlike the unbroken case where 2-point functions of fields with different conformal weights must vanish at separated points\footnote{They may however have contact terms at coincident points in certain cases \cite{Nakayama:2019mpz}.}.

\subsubsection{Scalar-tensor 2-point functions}
In the unbroken case, correlators of primaries with differing spin must vanish at separated points.  However, in the broken case there are allowed non-vanishing structures.  In the scalar-tensor case there is one such structure
\be \la \phi(X_1)T(X_2,Z_2)\ra\sim { \left( Z_2 \cdot V\,  X_1\cdot X_2-X_2\cdot V\, X_1\cdot Z_2 \right)^2 \over (X_1\cdot V)^{\Delta_1+2}(X_2\cdot V)^{\Delta_2+2}} F(\xi).\ee
This is the unique transverse structure with the correct scalings for both points.
Restricting to the CFT surface using \eqref{projectrelationsube}, \eqref{tfromdote} this becomes
\be \la \phi(x_1) t(x_2,z_2)\ra \sim { \left( {1\over 2}x_{12}^2\, v\cdot z_2+x_2\cdot v \, x_{12}\cdot z_2 \right)^2 \over  (x_1\cdot v) ^{\Delta_1+2}(x_2\cdot v)^{\Delta_2+2}}F(\xi) ,\ee  
where $x_{12}^\mu\equiv x_1^\mu-x_2^\mu$.
Stripping the auxiliary $z^\mu$ coordinates and using $v^\mu=(1,0,0,0)$ we obtain
\be \la \phi(x_1) t^{\mu\nu} (x_2)\ra \sim       {{1\over 4}x_{12}^4\delta^{(\mu}_t\delta^{\nu)_T}_t +(-t_{2})x_{12}^2 x_{12}^{(\mu} \delta^{\nu)_T}_t  +(-t_2)^2 x_{12}^{(\mu}  x_{12}^{\nu)_T}  \over  (-t_1)^{\Delta_1+2}(-t_2)^{\Delta_2+2}}F(\xi) .\ee    
where $(\ \ )_T$ means to take the symmetric traceless combination of the enclosed indices.

There are additional restrictions if $t^{\mu\nu}$ is conserved.  Setting the derivative to zero, $ \partial_\nu\la \phi(x_1) t^{\mu\nu} (x_2)\ra=0$, and then equating powers of $t_2$, yields an equation that forces the dimension of $t^{\mu\nu}$ to be the correct value for the stress tensor,
\be \Delta_2=4,\ee
and equating powers of $t_1$ provides a differential equation for $F$,
\be 6 (2 \xi -1) F(\xi )+2 (\xi -1) \xi  F'(\xi )=0,\ee
whose solution is
\be F(\xi)\sim \left(\xi(\xi-1)\right)^{-3}.\ee
Thus, for the stress tensor, the form of the scalar-vector 2-pt. function is completely fixed up to an overall constant.  Note that $F$ has singularities on the lightcone $\xi=0$ and on the surface $\xi=1$.   This possible transverse and traceless structure at separated points does not get realized in our holographic calculation.

\subsubsection{tensor-tensor 2-point functions}

The tensor-tensor 2pt function takes the form 
\be \la T(X_1,Z_1)T(X_2,Z_2)\ra\sim { F_1(\xi) H^2+F_2(\xi)HQ+F_3(\xi)Q^2 \over (X_1\cdot V) ^{\Delta_1}  (X_2\cdot V) ^{\Delta_2} },\ee
where 
\bea H & \equiv & { Z_1\cdot Z_2}-{X_1\cdot Z_2\, X_2\cdot Z_1\over X_1\cdot X_2\,}  , \\ 
Q & \equiv & \left( {X_1\cdot V \, X_2\cdot Z_1\over X_1\cdot X_2} +Z_1\cdot V\right)\left( {X_2\cdot V \, X_1\cdot Z_2\over X_1\cdot X_2} +Z_2\cdot V\right) \eea
are two transverse, symmetric, scale-invariant structures, and there are three undetermined functions $F_1$, $F_2$, $F_3$ of the cross ratio $\xi$.  Pulling back to the CFT surface using \eqref{projectrelationsube}, \eqref{tfromdote} this becomes
\be \la t(x_1,z_1)t(x_2,z_2)\ra\sim { F_1(\xi) h^2+F_2(\xi)hq+F_3(\xi)q^2 \over (-t_1) ^{\Delta_1}  (-t_2) ^{\Delta_2} },\ee
where
\bea  
h & \equiv &  z_1\cdot z_2-2{x_{12}\cdot z_2\, x_{12}\cdot z_1\over x_{12}^2\,}  , \\
q & \equiv & \left( 2  {(-t_1) \, x_{12}\cdot z_1\over x_{12}^2} -z_1^t \right)\left( -2{(-t_2) \, x_{12}\cdot z_2\over x_{12}^2} -z_2^t\right) .
\eea
 Imposing conservation fixes $\Delta_1=\Delta_2=4$, and also gives two equations that relate $F_1$, $F_2$, $F_3$, fixing two of them in terms of a third, so that there is only independent function of $\xi$.

\renewcommand{\em}{}
\bibliographystyle{utphys}
\addcontentsline{toc}{section}{References}
\bibliography{2pointPCU-arxiv}

\end{document}